\documentclass[onecolumn]{aastex7}

\usepackage{amsmath}
\usepackage{graphicx}
\usepackage{natbib}
\usepackage{xspace}
\usepackage{xcolor}
\usepackage{hyperref}
\usepackage{makecell}

\newcommand{\mcl}{M_{\rm cl}\xspace}
\newcommand{\mmin}{m_{\rm\star,\,min}\xspace}
\newcommand{\mmax}{m_{\rm\star,\,max}\xspace}
\newcommand{\msun}{\ensuremath{\mathrm{M}_\odot}\xspace}
\newcommand{\imf}{\texttt{imf}\xspace}
\newcommand{\mtot}{\mathrm{M}_{\rm tot}\xspace}

\begin{document}

\title{The IMF package: a toolkit implementing mass functions and statistical tools to analyze them}

\author[0009-0001-8880-6951]{Theo Richardson}
\affiliation{Department of Astronomy, University of Florida, PO Box 112055, Gainesville, FL, USA}
\email{terichard57@gmail.com}

\author[0000-0001-6431-9633]{Adam Ginsburg}
\affiliation{Department of Astronomy, University of Florida, PO Box 112055, Gainesville, FL, USA}
\email{adamginsburg@ufl.edu}

\author[0000-0003-2644-135X]{Sergey E. Koposov}
\affiliation{Institute for Astronomy, University of Edinburgh, Royal Observatory, Blackford Hill, Edinburgh EH9 3HJ, UK}
\affiliation{Institute of Astronomy, University of Cambridge, Madingley Road, Cambridge CB3 0HA, UK}
\email{skoposov@ed.ac.uk}

\begin{abstract}
Mass functions are used in all areas of astrophysics. The stellar initial mass function (IMF), in particular, plays a central role in modeling stellar populations in galaxies. However, few dedicated tools for working directly with the IMF and its precursor functions are widely available. We present the \imf package, a Python library integrated into the wider scientific Python ecosystem that implements common and variant forms of mass functions, especially the IMF and its pre- and protostellar equivalents, as probability distribution functions based on SciPy's statistics architecture. This package enables the performance of operations such as sampling and integration on a wide array of highly customizable mass functions. \imf is publicly available on the Python package index \texttt{pypi} under the project name \texttt{initial\_mass\_function}.

\end{abstract}


\section{Motivation} \label{sec:intro}
Mass functions have been a critical tool across astrophysics over the last half century. The probability of producing an object of a given mass is a natural output of formation theory for objects across mass scales (e.g. stars, star clusters, galaxies, dark matter halos) and is therefore a widely used point of comparison between observation and theory \citep[e.g.][]{jenkins2001,lada2003,somerville2015}. A particularly prominent example of the utility of mass functions is the stellar initial mass function (IMF), the mass distribution of newly formed stars. The appearance, age, and evolution of a star are all governed by its mass, so knowing how a star starts its life is essential for modeling the formation and evolution of star clusters and galaxies and the distribution of planetary systems.

Many models of the IMF have been proposed and utilized throughout the literature. The most prominent are those of \citet{salpeter1955}, a pure power law with exponent $\Gamma=1.3$ favored for its simplicity and commonly used in whole-galaxy models, \citet{kroupa2001}, the favored broken power-law distribution, and \citet{chabrier2003a}, which combines a lognormal with a high-mass power law; variants of these favored forms and alternate distributions have been proposed or measured throughout the life of the IMF as a concept.

The principal use for these IMF models is as a component of stellar population models. The foremost examples of these models are simple stellar populations, often abbreviated as SSPs, which contribute to stellar population synthesis (SPS) techniques and are therefore a cornerstone of modern measurements of galaxy properties \citep{conroy2013}.
Beyond the context of population synthesis, the IMF is one of the principal observable quantities in star formation theory 
and is a key component in calculations of population-scale quantities, such as expected supernova rates.

However, despite this importance, few publicly available tools exist to work with these IMF models. The tools that do exist are either larger packages that do not focus directly on IMF sampling, instead making it one component of a larger enterprise \citep[e.g.][]{krumholz2015}, or tools that focus primarily on particular forms and conceptions of the IMF for specific purposes \citep[e.g.][]{yan2017,gjergo2026b}. While some forms of the IMF are trivial to implement, others are not, and applications of the IMF that involve sampling can be challenging.

Further, uncertainty remains regarding the origin of the IMF. There are many competing models for the formation of the IMF. These models make predictions about the shape of precursors to the IMF, including the core mass function (CMF) and the protostellar mass function (PMF). However, these models have not been widely used or tested, in large part because implementing them is generally more challenging than common IMF forms; they often contain time-dependent terms and exhibit a higher level of mathematical complexity.

We present the \imf package, which enables straightforward sampling of mass functions across multiple functional forms and according to multiple sampling methods, providing a highly flexible set of tools to facilitate mass-function-dependent modeling. \imf is a lightweight and modular mass function sampling package that is integrated with the wider ecosystem of software used for astronomy. It retains the customizability and robustness of prominent SPS codes \citep[e.g.][]{krumholz2015} and remains general enough to easily use outside the context of the stellar IMF, including present-day mass functions or nonstellar use cases for mass function models. Furthermore, \imf allows access into the pre- and protostellar phases of stellar evolution, an area beyond the reach of previous IMF sampling codes.

The package is publicly available and is distributed on the python package index as \texttt{initial\_mass\_function}.


\section{Mass Functions}\label{sec:mfs}
\imf provides implementations of multiple commonly used mass functions, each an instance of an overarching \texttt{MassFunction} class. These \texttt{MassFunction}s are built out of \imf's \texttt{Distributions}, which are basic customizable statistical models that enable the functions to be integrated and randomly sampled. \texttt{Distributions} are either wrappers for corresponding functions within \texttt{scipy}'s \texttt{stats} module or custom functions.
Within the custom functions, integration is performed with \texttt{scipy.integrate.quad} and interpolation with \texttt{scipy.interpolate.PchipInterpolator}\footnote{\texttt{PchipInterpolator} uses the PCHIP algorithm \citep{fritsch1984} to interpolate using cubic splines.}. Interpolation is applied in cases where the mass function is particularly expensive to calculate or cannot be sampled otherwise; usages of interpolation will be noted throughout the paper and are similarly identified in the package documentation. 
We choose PCHIP interpolation because it is faithful to the base functions and it preserves monotonicity, both of which are important for the use cases in \imf. Interpolated functions generally depart from the true values; however, numerical testing indicates that the expected error will remain within $\sim.1\%$ of true value for sufficiently well-sampled functions. By default, interpolated functions are sampled at 200 points, but the number of points is a tunable parameter. Functions that rely on interpolation cannot accept nonfinite $\mmin$ or $\mmax$; if a function relies on interpolation, it will default to finite mass bounds.

The remainder of this section describes the implementation of all \texttt{MassFunction}s. Each \texttt{MassFunction} provides access to the probability distribution function (PDF) $\xi(m)$, the cumulative distribution function (CDF) $\Xi(m) \equiv\int_{\mmin}^m \xi(m)dm$, and the mass-weighted PDF $m\times \xi(m)$ of its underlying \texttt{Distribution}, where the implemented mass function is given by the PDF. IMFs are also commonly displayed in a logarithmic form $\xi(\ln m)$, which is equivalent to the mass-weighted PDF (by definition of a PDF, $\int_{\rm min}^{\rm max} \xi(\ln m)d\ln m=\int_{\rm min}^{\rm max} \xi(m)dm=1$; because $d\ln m=dm/m$, $\xi(\ln m)$ must equal $m\times \xi(m)$.)

\begin{figure*}
    \centering
    \includegraphics[width=0.65\textwidth]{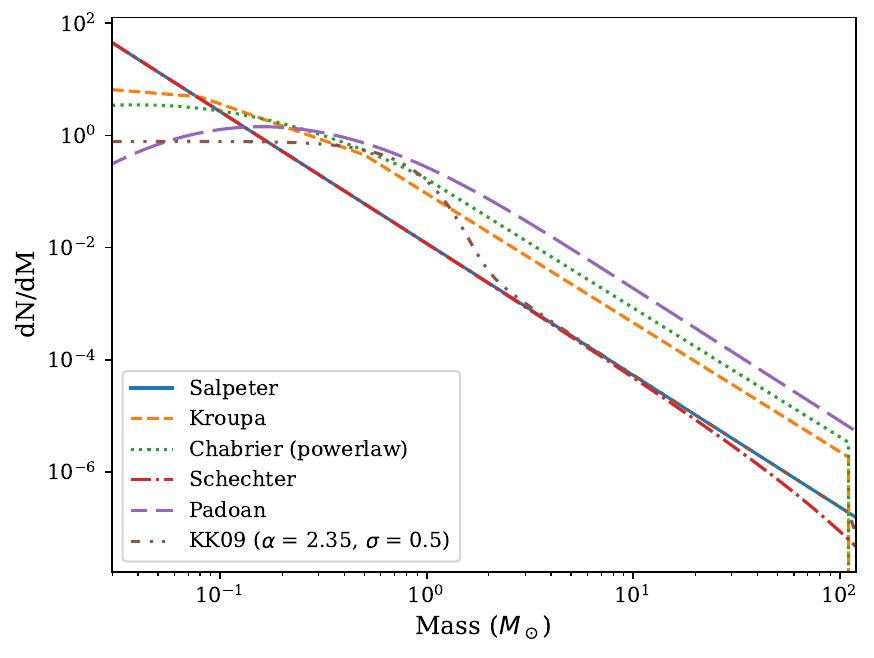}
    \caption{A sample of mass functions implemented in \imf. All plotted functions span a mass range of 0.03-120 $\msun$. Each IMF uses its default parameters where possible. For the error-convolved power law (\S\ref{sec:mfs.5}), which has no defaults, parameters are included in the legend.}
    \label{fig:imfs}
\end{figure*}

Figure \ref{fig:imfs} displays a sample of available mass functions, plotted over a typical stellar mass range for visual comparison.
These forms can be supplied with alternate parameter values; as an example, a basic power law \texttt{MassFunction} may have its slope changed to model IMFs that are more ``bottom-/top-heavy" \citep[e.g.][]{conroy2012,schneider2018}. Further, \imf provides the infrastructure to create new \texttt{Distribution}s, either by combining implemented ones using the \texttt{CompositeDistribution} class or subclassing \texttt{Distribution}.

The mass functions described here are all system mass functions, i.e. in the context of the IMF they describe the mass distribution of stellar systems (which may be multiples) as opposed to the distribution of individual stellar masses, which treats every star as its own unique constituent. The actual stellar IMF resulting from subdivision into multiple systems is expected to have a slightly steeper high-mass slope than the system IMF across the range of $\alpha$ measured in the literature \citep{rosen2026}. \imf also provides functions to create populations that sample down to the stellar level; see Section \ref{sec:sampling.3.1}.

\subsection{Power law}\label{sec:mfs.1}
The original parameterization of the IMF in \citet{salpeter1955} is a simple power law, following the form:
\begin{equation}
    \xi(m)\propto m^{-\alpha}.
    \label{eq:powerlaw}
\end{equation}
The canonical function has lower- and upper-mass cutoffs, typically assumed to be 1 and 100 \msun, respectively, and a slope $\alpha=2.35$ ($\Gamma=1.35$). \imf implements a \texttt{Salpeter} mass function based on a \texttt{PowerLaw} distribution. This implementation is similar to the canonical function; however, $\mmin=0.3\msun$ and $\mmax=120\msun$ by default for greater consistency with the other models.

\subsection{Broken power law}\label{sec:mfs.2}
Multiple works have modeled the IMF as a broken power law with segments spanning particular ranges of stellar and substellar mass. The most prominent of these comes from \citet{kroupa2001}, while a recent census of nearby objects \citep{kirkpatrick2024} provides a broken power law that more accurately captures the local substellar population. \imf provides a \texttt{BrokenPowerLaw} class to encompass IMFs of this type. A \texttt{BrokenPowerLaw} with $n$ segments creates a list of multiple \texttt{PowerLaw} distributions with slopes $p_i,\,i\in(1, 2...n)$ that meet at break points $b_j,\,j\in(1,2...n-1)$; both slopes and break points must be provided. Each \texttt{PowerLaw} is scaled appropriately to ensure the overall distribution is continuous and normalized. Calling the resulting function identifies the appropriate \texttt{PowerLaw}(s) to access using the provided stellar mass.

\imf includes implementations of both the \citet{kroupa2001} and \citet{kirkpatrick2024} broken power-law mass functions.
The default \texttt{Kroupa} IMF has the following parameters: $p_1 = 0.3$ up to $b_1 = 0.08 \msun$, $p_2 = 1.3$ up to $b_2 = 0.5 \msun$, and $p_3 = 2.3$ above $b_2$. This corresponds to Equation 2 of \citet{kroupa2001} and is consistent with later restatements \citep{kroupa2013,kroupa2026}. The \texttt{Kirkpatrick2024} IMF has the following parameters: $p_1 = 0.6$ up to $b_1 = 0.05 \msun$, $p_2 = 0.25$ up to $b_2 = 0.22 \msun$, $p_3 = 1.3$ up to $b_3 = 0.55 \msun$, and $p_4 = 2.3$ above $b_3$. Both functions have $\mmin=0.03\,\msun$ and $\mmax=120\,\msun$ by default.

\subsection{Lognormal}\label{sec:mfs.3}
\citet{chabrier2003a,chabrier2003b} creates an IMF model based on mass measurements of stars in the Galactic disk; the resulting function has a lognormal shape for stars with masses $\lesssim 1\msun$ and a power-law shape for more massive stars. The lognormal component follows this general form:
\begin{equation}
    \xi(m)=\frac{A}{m}\times\exp\left[-\frac{1}{2}\times\left(\frac{\log(m)-\log(m_{\rm\star,\,mean})}{\sigma}\right)^2\right]
    \label{eq:lognormal}
\end{equation}
for scale factor $A$, peak of the distribution $m_{\rm\star,\,mean}$, and width shape parameter $\sigma$.

\imf implements two variants of the Chabrier IMF: one that is completely lognormal (\texttt{ChabrierLogNormal}) and one that combines the lognormal with a power law (\texttt{ChabrierPowerLaw}). The latter is based on \imf's \texttt{CompositeDistribution}, which joins multiple \texttt{Distribution}s together in a manner similar to \texttt{BrokenPowerLaw}. By default, the components are joined at 1 $\msun$ and the power law has $\alpha=2.3$ in accordance with the canonical form, but both the transition point and power-law slope can be changed. The default implementations of both functions adopt the lognormal shape parameters corresponding to the parameterized system mass function (Equation 18 in \citet{chabrier2003a}/Equation 2 in \citet{chabrier2003b}); hence, $A=0.086$, $m_{\rm\star,\,mean}=0.22$, and $\sigma=0.57$\footnote{\citet{chabrier2005} provides alternate numbers based on a revised disk stellar luminosity function; these are $A=0.076$, $m_{\rm\star,\,mean}=0.25$, and $\sigma=0.55$.}; these parameters may also be altered.

\subsection{Tapered power law}\label{sec:mfs.4}
The mass function of large-scale objects (e.g. star clusters, galaxies, dark matter halos) is often modeled as a power law with an exponentially tapered high-mass end, following the formalism of \citet{press1974}:
\begin{equation}
    \xi(m)\propto m^{-\alpha}e^{-m/m_{\rm c}}.
    \label{eq:taper}
\end{equation}
\imf implements such a form in the \texttt{Schechter} \texttt{MassFunction}.
By default, the underlying power law will have the same slope ($\alpha$) as the Salpeter IMF, while the characteristic mass $m_{\rm c}$ for exponential tapering defaults to 100 $\msun$. \imf also provides a \texttt{ModifiedSchechter} function, which tapers both the low- and high-mass ends of the underlying power law:
\begin{equation}
    \xi(m)\propto m^{-\alpha}e^{-m_{\rm l}/m}e^{-m/m_{\rm u}}.
    \label{eq:modtaper}
\end{equation}
$m_{\rm u}$ retains the default of 100 $\msun$, while the characteristic mass for the low-mass end $m_{\rm l}$ defaults to 0.5 $\msun$.

In keeping with other implemented functions, the default properties for the \texttt{Schechter} function position it as a stellar IMF in place of more traditional uses. While most canonical forms for the IMF are approximately pure power laws over some fraction of the stellar mass range, a tapered mass function results in a smooth decline in the expected number of objects beyond a particular mass, providing an alternative to the traditional assumption of hard stellar mass limits. However, because all properties of these Schechter-like functions are tunable, they can be easily applied in a more typical fashion (e.g. sampling star cluster masses as done in Section \ref{sec:sampling.3.2}).

Neither \texttt{Schechter} nor \texttt{ModifiedSchechter} have CDFs with well-defined inverses, which is a necessity for random sampling as handled by \imf (see \S\ref{sec:sampling.1}); consequently, this function relies on interpolation for sampling.

\subsection{Error-convolved power law}\label{sec:mfs.5}
\citet[][KK09]{koen2009} define a probability distribution for the convolution of a power law with a Gaussian, intended to simulate data that follows a power law but is contaminated by measurement errors.
\imf implements this error-convolved power law. The mass function that follows the PDF for this error-convolved power law can be written as:
\begin{equation}
    \xi(m)=\frac{1}{\sigma\sqrt{2\pi}}\times\frac{\alpha}{m_{\rm min}^{-\alpha}-m_{\rm max}^{-\alpha}}\int_{m_{\rm min}}^{m_{\rm max}}x^{-(\alpha+1)}{\rm exp}\left[-\frac{1}{2}\left(\frac{m-x}{\sigma}\right)^2\right]dx.
    \label{eq:koenpdf}
\end{equation}
The corresponding CDF is:
\begin{equation}
    \Xi(m)=\Phi\left(\frac{m-m_{\rm max}}{\sigma}\right)+\frac{1}{\sigma\sqrt{2\pi}}\times\frac{\alpha}{m_{\rm min}^{-\alpha}-m_{\rm max}^{-\alpha}}\int_{m_{\rm min}}^{m_{\rm max}}\left(m_{\rm min}^{-\alpha}-x^{-\alpha}\right){\rm exp}\left[-\frac{1}{2}\left(\frac{m-x}{\sigma}\right)^2\right]dx
    \label{eq:koencdf}
\end{equation}
where $\sigma$ comes from the Gaussian component and corresponds to measurement error\footnote{Note that these equations generally require the imposition of a lower and upper limit on the variable in order to evaluate. For a mass function, these emerge naturally as the lower and upper limits on mass.}.
Note that these equations assume that $\sigma$ is a constant, which is generally not a good assumption for stellar mass measurements; in practice, this functional form has largely been invoked for objects on larger scales (e.g. molecular clouds, galaxies, etc.). \imf includes this form to provide access to the shape of the distribution.
$\Phi$ is the standard normal CDF:
\begin{equation}
    \Phi(x)=\frac{1}{\sqrt{2\pi}}\int_{-\infty}^x e^{-t^2/2}dt.
\end{equation}
Since the KK09 distributions are chiefly meant to model measurements with uncertainties, there is no ``canonical" KK09-esque mass function. \imf therefore does not include default values, meaning an $\alpha$, $\sigma$, and mass range must be supplied to create an instance. 

These equations do not have simple analytic forms. \imf solves them through numeric integration. However, the integrands of Equations \eqref{eq:koenpdf} and \eqref{eq:koencdf} span many orders of magnitude when evaluated over the full stellar mass range. This can result in inaccuracies, particularly near the endpoints of the integral. \imf includes two implementations of the KK09 functions, \texttt{KoenConvolvedPowerLaw} and \texttt{SpotKoenConvolvedPowerLaw}, which solve this problem in distinct ways. \texttt{KoenConvolvedPowerLaw} evaluates the PDF and CDF at a set of points across the mass range of the underlying function and interpolates between the resulting values. In contrast, \texttt{SpotKoenConvolvedPowerLaw} does all of its calculations ``on-the-spot", meaning that each call to the PDF or CDF actually evaluates the function.
Both versions perform the integration necessary to evaluate the function by splitting the domain of integration into several subdomains. These subdomains are generally evenly linearly spaced, but switch to evenly log-spaced near the edges of the domain, with the linear space covered by each step decreasing nearing the respective endpoint. This semiadaptive approach is better able to capture the small values attained by the integrand near the endpoints, leading to increased accuracy. However, because \texttt{quad} must be called for each subdomain, it also results in an increased time cost.
\begin{deluxetable*}{c c c}
    \tabletypesize{\small}
    \tablewidth{\textwidth}
    \tablehead{\colhead{Class} & \colhead{Pros} & \colhead{Cons}}
    \tablecaption{Comparison between \imf's implementations of the \citet{koen2009} error-convolved power law.}\label{tab:koen}
    \startdata
    \texttt{KoenConvolvedPowerLaw} & \makecell{Quick evaluation ($\approx0.01$ ms),\\random sampling} & \makecell{Slow instantiation ($\approx10$ s),\\slight inaccuracy} \\
    \texttt{SpotKoenConvolvedPowerLaw} & \makecell{Quick instantiation ($\approx0.01$ s),\\good accuracy} & \makecell{Slow evaluation ($\approx0.04$ s),\\no random sampling}
    \enddata
\end{deluxetable*}

The motivation for multiple implementations of the KK09 power law stems from the fact that both versions have consequential drawbacks. 
Table \ref{tab:koen} summarizes their pros and cons. 
Each version suffers from the time cost from the integration method, causing slow instantiation for \texttt{KoenConvolvedPowerLaw} and slow evaluation for \texttt{SpotKoenConvolvedPowerLaw}, which are the points where the integrals in Equations \eqref{eq:koenpdf} and \eqref{eq:koencdf} are performed for each class.
\texttt{KoenConvolvedPowerLaw}, because it interpolates, has values that are slightly inaccurate.
Meanwhile, \texttt{SpotKoenConvolvedPowerLaw} cannot be randomly sampled. The infrastructure of random sampling in \imf relies on a \texttt{MassFunction}'s underlying \texttt{Distribution}, which \texttt{SpotKoenConvolvedPowerLaw} does not have due to its on-the-spot calculations. Moreover, as random sampling relies specifically on the inverse of the CDF (see \S\ref{sec:sampling.1}), and because Equation \ref{eq:koencdf} involves a definite integral, it cannot be inverted. Consequently, no analytic form exists for the inverse CDF of a KK09 power law; unlike the CDF, the inverse CDF cannot be evaluated on the spot. As a result, random sampling is impossible without constructing a lookup table for CDF values, which is already covered by \texttt{KoenConvolvedPowerLaw}.

\begin{samepage}
\subsection{The IMF derived from the Core Mass Function assuming turbulent fragmentation}
\nopagebreak
\label{sec:mfs.6}
\nopagebreak
\citet{padoan1997} and \citet{padoan2002} derive a form for the IMF that emerges from their model of dense core formation from shocked gas within a supersonically turbulent molecular cloud. Mathematically, this form is:
\begin{equation}
    \xi(m)\propto m^{-4/(4-b)}\times\int_0^m p(m_{\rm J})dm_{\rm J}
    \label{eq:padoan}
\end{equation}
where $p(m_{\rm J})$, the Jeans mass distribution, is
\begin{equation}
    p(m_{\rm J})=\frac{2m_{\rm J,\,0}^2}{\sqrt{2\pi}\sigma}m_{\rm J}^{-3}\times \exp\left[-\frac{1}{2}\left(\frac{2\ln m_{\rm J}/m_{\rm J,\,0} - \sigma^2/2}{\sigma}\right)\right].
    \label{eq:pmj}
\end{equation}
$\sigma$ is the standard deviation of the lognormal distribution of density in a turbulent flow and $m_{\rm J,\,0}$ is the Jeans mass at the average density, defined as $1.2\,\msun(T/10\,{\rm K})^{3/2}(n_0/1000\,{\rm cm}^3)^{-1/2}$. (Note that both Equations \eqref{eq:padoan} and \eqref{eq:pmj} are stated here in their linear, non-mass-weighted forms.) This IMF form is dependent on the properties of the parent cloud and the turbulence within it, but qualitatively resembles a Chabrier IMF; it generally emerges as a power law above $\approx1\,\msun$ with a peak between 0.2 and 0.6 $\msun$.
\end{samepage}

\imf implements this IMF form in the \texttt{PadoanTF MassFunction}. \texttt{PadoanTF} is handled similarly to error-convolved power laws (\S\ref{sec:mfs.5}) in that the IMF is calculated at a set of points on instantiation and further calls interpolate between these precalculated values in order to enable random sampling. The tunable parameters unique to \texttt{PadoanTF} set the cloud and turbulence properties: $b$ (slope of the turbulence power spectrum, defaults to 1.8), $T_0$ (mean cloud temperature, defaults to 10 K), $n_0$ (mean cloud number density, defaults to 500 cm$^{-3}$), and $\sigma$. \texttt{PadoanTF} can also be provided with $\mathcal{M}$--the rms Mach number of the turbulent flow--as an alternative to $\sigma$, which will then be calculated using $\sigma^2=\ln(1+(\mathcal{M}/2)^2)$ following Equation 9 of \citet{padoan2002}. If both $\sigma$ and $\mathcal{M}$ are provided, only $\sigma$ will be used in IMF calculation. $\mathcal{M}$ defaults to 10. Like error-convolved power laws (\S\ref{sec:mfs.5}), \texttt{PadoanTF} relies on interpolation both for evaluation and sampling due to the complexity of the underlying functions.

It should be noted that this function technically describes the mass distribution of cores--dense condensations of gas and dust that supply mass to forming stars--instead of a true stellar IMF. \citet{padoan2002} adopts the framing of a direct correspondence between the masses of cores and resulting stars such that the two are roughly equivalent. For the purpose of representing the morphology of the mass function in the package, this distinction is therefore elided here. However, \imf does also explicitly provide mass functions for cores; see Section \ref{sec:func.1.2}.

\section{Sampling Techniques}\label{sec:sampling}
\begin{figure}
    \centering
    \includegraphics[width=0.32\textwidth]{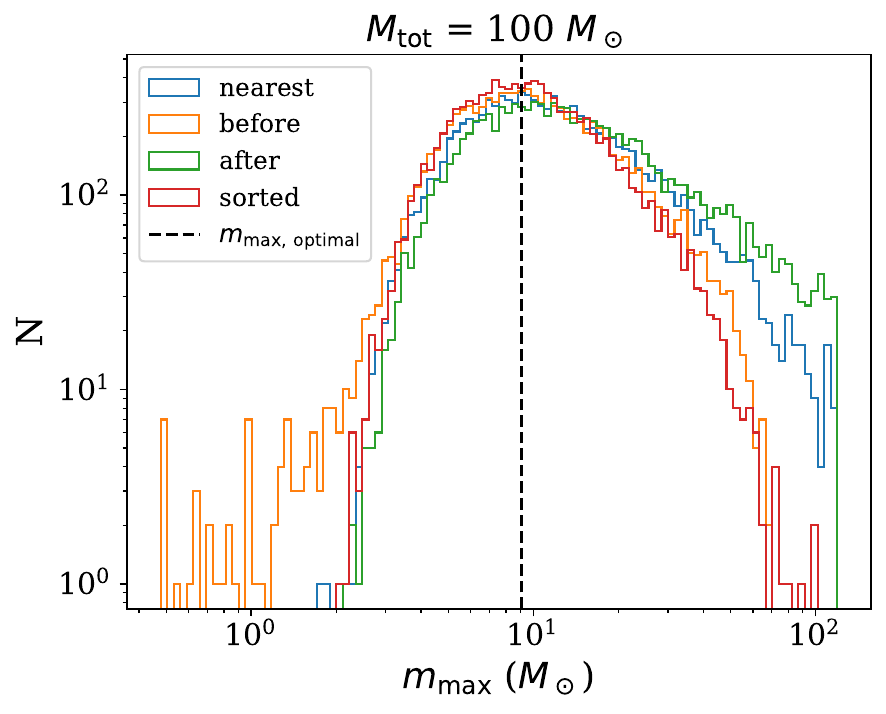}
    \includegraphics[width=0.32\textwidth]{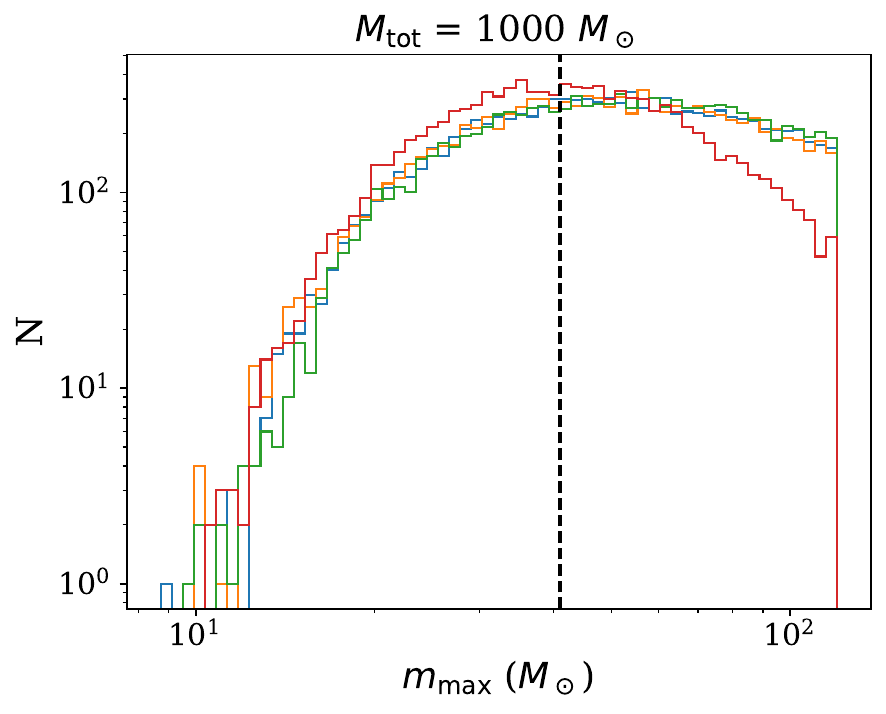}
    \includegraphics[width=0.32\textwidth]{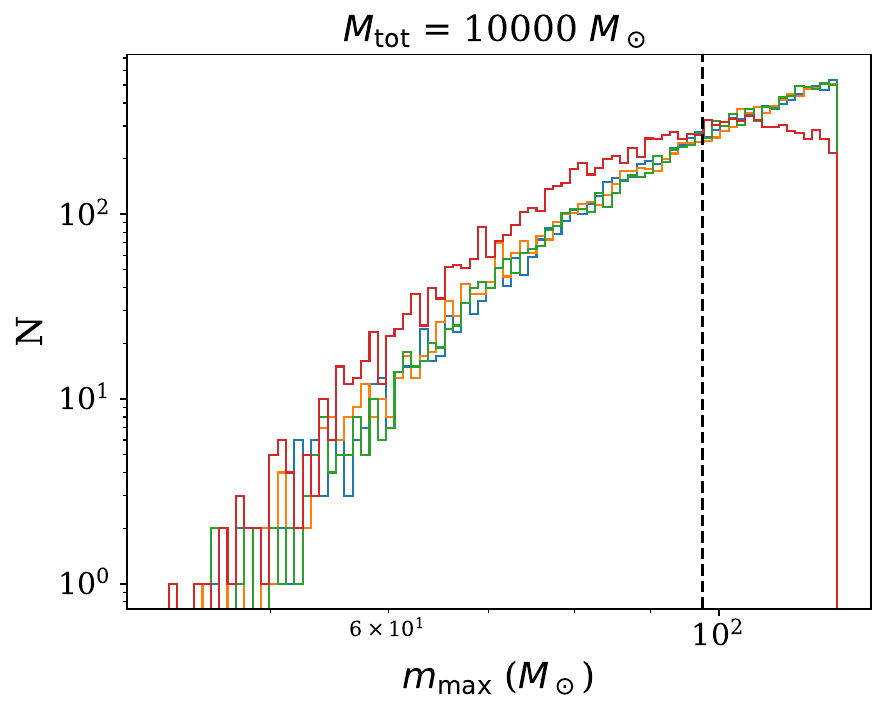}
    \caption{The most massive star systems in clusters of 10$^2$/10$^3$/10$^4$ $\msun$ (\textit{left/middle/right}) sampled from a Kroupa IMF according to each of \imf's methods. Random sampling (\S\ref{sec:sampling.1}) is shown as histograms that plot the distribution of the most massive system taken from 10$^4$ cluster realizations at each mass. Optimal sampling (\S\ref{sec:sampling.2}) is shown as a vertical line marking the most massive system possible given the cluster mass and default Kroupa stellar mass range.}
    \label{fig:sampling}
\end{figure}

\imf's central functionality is sampling masses from modeled mass functions. This sampling is handled through the functions \texttt{sample\_mass} and \texttt{sample\_number}, which create a collection of masses meeting either a provided mass budget or number of members, respectively. \imf has multiple ways to perform this sampling, as the method used to sample has been demonstrated to exert nontrivial systematic effects on resulting populations, particularly for those with lower mass budgets. In this section, we detail these methods.
Figure \ref{fig:sampling} demonstrates the impact of the sampling method on resulting populations as characterized through their most massive stars, similar to \citet{krumholz2015}.

\subsection{Random sampling}\label{sec:sampling.1}
\imf allows random sampling of masses from a mass function, which is common practice for SSPs and similar models. Random sampling can be done either by providing a mass budget or a number of draws. Random sampling by mass works by drawing $M_{\rm rem}/\left<M\right>$ masses from a mass function (where $M_{\rm rem}$ is $\mtot-M_{\rm pop,\,current}$, the total remaining mass not contained in sampled masses, and $\left<M\right>$ is the function's mass-weighted average) repeatedly until the provided mass budget is exceeded. \imf provides a ``tolerance" keyword that shifts the threshold beyond which sampling stops to $\mtot+M_{\rm tolerance}$. The tolerance may be positive or negative.

Masses are drawn by calling the \texttt{rvs} method of the \texttt{Distribution} underlying the \texttt{MassFunction}\footnote{\texttt{rvs} originates from \texttt{scipy}'s implementation of random sampling for its statistical distributions. \imf's \texttt{Distribution}s use the same syntax for consistency.}, which randomly samples a uniform continuous distribution in the interval $[0,1)$ and maps the resulting values to mass space using the inverse CDF--otherwise known as the percent-point function (PPF)--of the \texttt{Distribution}. This sampling method is natively implemented in \texttt{Distribution}s that wrap \texttt{scipy.stats}, and custom \texttt{Distribution}s are structured to work identically.

Once the mass budget is exhausted, stars sampled in the final draw are either kept or discarded depending on the active sampling stop criterion, and the resulting cluster is returned. Said criterion corresponds to the algorithm used to handle the final draw from the IMF, which may be any of the following:
\begin{itemize}
    \item ``nearest": Include all stars drawn from an IMF (in drawing order) that bring the cumulative mass of the cluster closest to $\mtot$. Sometimes exceeds $\mtot$.
    \item ``before": Include all stars drawn from an IMF (in drawing order) with cumulative mass $<\mtot$. Never exceeds $\mtot$.
    \item ``after": Include all stars drawn from an IMF (in drawing order) with cumulative mass $<\mtot$, and also the next star. Always exceeds $\mtot$.
    \item ``sorted": Sort the stars by mass in increasing order, then include or exclude stars based on the ``nearest" criterion such that only the most massive sampled stars are subject to exclusion.
\end{itemize}
Each of these algorithms are adapted from \citet{krumholz2015}, although sorted sampling originates from \citet{weidner2006}. \imf defaults to ``nearest". For an example of ``nearest": We intend to create a star cluster with $\mtot=1000$ $\msun$, and we have sampled to a total of 950 $\msun$. The next three sampled star systems have masses (0.2, 45, 10) $\msun$. $995.2 = 950 + 0.2 + 45 \msun$  is closer to $\mtot$ than $1005.2 = 950 + 0.2 + 45 + 10$, so the first two are included and none after. If the next three sampled systems were instead (0.2, 10, 45) $\msun$, all three would be included because 1005.2 would be closer to 1000 than $960.2=950+0.2+10$.

Random sampling by number is comparatively simple; the number of requested draws is passed directly to the \texttt{Distribution}'s \texttt{rvs} method, which is designed to accept a number of samples as input. The stop criterion therefore has no impact on sampling a number of masses, as it exclusively deals with managing an available mass budget.

\subsection{Optimal sampling}\label{sec:sampling.2}
In addition to the random sampling methods described in the previous section, \imf allows for the creation of populations that ``optimally" sample mass functions. The populations created by optimal sampling perfectly reproduce the shape of the underlying mass function and fully use the available mass budget. Unlike the methods in Section \ref{sec:sampling.1}, optimal sampling is deterministic, i.e. for a given mass budget and mass function there is only one possible ``optimal" population.
Optimal sampling is therefore a distinct paradigm, which has been applied across multiple size scales \citep[e.g.][]{kroupa2013,schulz2015,yan2017} and is argued to occur in the star formation process as a consequence of maximally entropic gravitational fragmentation \citep{gjergo2026a}.

Optimal sampling is implemented following \citet{kroupa2026}, which summarizes the original prescription of \citet{kroupa2013} as modified by \citet{schulz2015}. Effectively, this procedure divides integrals reproducing the total mass and number of members of a population into a series of integrals between monotonically decreasing bounds $b_0,\,b_1,\,...\,b_n$ which each satisfy $\int_{b_i}^{b_{i-1}}\xi(m)dm=1$, i.e. it finds the points in the mass function describing the population with one member between them. The mass of each member is then the integral of the mass-weighted mass function, i.e. $m_i =\int_{b_i}^{b_{i-1}}m\xi(m)dm$. This process is necessarily iterative, as the boundaries must be found in succession by integrating down the function. A total mass budget is necessary for optimal sampling as conceived by \citet{kroupa2013} and therefore as implemented in \imf, meaning that all optimal sampling eventually uses the \texttt{sample\_mass} function. Optimal sampling can also be called from \texttt{sample\_number}; \imf will translate this to a mass budget using the expectation value of the provided mass function (i.e. a total mass of $N\times\left<M\right>$). The resulting population will generally have a different number of members than requested, but the difference will be small.

\imf's algorithm is as follows: after providing a total mass budget and mass function, \imf calculates the most massive member of the population and solves for successive, less massive members until the mass budget is completely used. Finding the mass of the most massive member, $m_{\rm (most\,massive)}$, requires solving the following system of equations in order to properly scale the function:
\begin{align}
    k(m)&=\left(\int_{m}^{m_{\rm max}}\xi(m')dm'\right)^{-1}\label{eq:opt_k}\\
    \mcl(m)= m&+k(m)\times\int_{m_{\rm min}}^{m}m'\xi(m')dm'.
    \label{eq:opt_mcl}
\end{align}
These are solved by using root finding on Equation \eqref{eq:opt_mcl}. Root finding is done via \texttt{scipy}'s \texttt{root\_scalar} function.
Once the most massive member in the optimally sampled population is found and the normalization is set accordingly, the masses of new members are calculated using
\begin{equation}
    m_i = k(m_{\rm (most\,massive)})\times\int_{b_i}^{b_{i-1}}m\xi(m)dm,
    \label{eq:opt_nextstar}
\end{equation}
where $b_{i-1}$ and $b_i$ enforce the one-member condition (i.e. $\int_{b_i}^{b_{i-1}}\xi(m)dm=1$) for each successive $i$. The sequence of bounds begins with $b_0=\mmax$ and $b_1=m_{\rm(most\,massive)}$, which meet the one-member condition thanks to solving Equation \eqref{eq:opt_k}. New members are sampled until the remaining mass budget is consumed. Since the mass of each successive member is always less than the previous member, all root finding is bracketed.

Note that the iterative nature of the optimal sampling process means that a closing condition is required; in turn, this necessitates a positive nonzero lower bound on mass. This lower limit is taken to be either the minimum mass of the provided mass function or the tolerance provided to \texttt{sample\_mass} if it is positive. The minimum mass is preferred; however, if it is zero, the tolerance will be used instead. \imf will not make an optimally sampled population if both values are zero or negative. The existence of a lower limit for optimal sampling means that the provided mass budget will not actually be used completely; however, it will be allocated down to the provided lower mass limit (i.e. optimal sampling will make as many members as possible given the provided lower bound).

\subsection{Specific use cases}
This section describes sampling techniques developed for particular science applications; these introduce additional algorithmic complexity, but generally use the machinery described in Sections \ref{sec:sampling.1} and \ref{sec:sampling.2} and share a module in the package.

\subsubsection{Star clusters}\label{sec:sampling.3.1}
\imf provides a function that creates multiple-aware star clusters, i.e. it extends masses sampled from a system IMF down to the stellar level. This function samples a population of systems, assigns each system a multiplicity, determines the mass ratios between components, and uses this information to convert the system population into a stellar one. 
Multiplicity is assigned randomly to each system based on the multiplicity fractions of observed systems compiled by \citet{offner2023};
appropriate probabilities are determined by interpolating between these fractions at the system mass. 
Systems may be singles, binaries, or triples (as a general stand-in for higher-order multiples). 
Once the multiplicity of each system is determined, the mass ratios for all nonprimary members ($q\equiv m/m_{\rm primary}$) are randomly drawn from a uniform distribution and used to calculate the masses of each star within the system. The ability to preserve the association between members of a multiple system (i.e. to group the masses of stars in systems together) can be toggled.

\subsubsection{Integrated galaxy IMFs}\label{sec:sampling.3.2}
The integrated galaxy IMF (IGIMF) is conceived as the distribution describing all stars formed across a galaxy, which emerges from the sum of IMFs of star clusters formed within the galaxy \citep{kroupa2003,yan2017}. \imf provides a function to calculate IGIMFs. In keeping with existing implementations of the IGIMF theory, \imf samples either a total mass or number of star clusters and then samples star systems from the resulting mass reservoirs. Cluster masses are drawn from a Schechter function, which is a reasonable approximation to the mass function of star clusters in local galaxies \citep[e.g.][]{johnson2017,wainer2022}. By default, this function will have $\alpha=2$ and $m_c=8.5\times10^3\,\msun$ in accordance with the mass function measured in M31 by \citet{johnson2017} and will be defined between $10^2$ and $10^6\,\msun$, the observed mass range for Galactic star clusters \citep{portegieszwart2010}, but all of these properties are tunable. The mass function and sampling method used to sample star systems are likewise tunable. The resulting population will provide the IGIMF, with the minor caveats that the IGIMF theory used to construct this approach builds in the assumption that all star formation happens in clusters and that sampling and storage of the IGIMF can potentially be expensive in time and memory (see Section \ref{sec:perf}).

\section{Performance}\label{sec:perf}
In this section, we characterize the performance of \imf, focusing on the speed and memory usage of sampling. In order to provide useful and standardized performance data, we sample star clusters\footnote{Since this is intended as a test of the basic operations, these clusters are not extended to stellar masses.} across the Galactic cluster mass range ($10^2-10^6\,\msun$) according to all of our implemented mass functions (\S\ref{sec:mfs}) using each of our techniques (\S\ref{sec:sampling}).
All mass functions have their default parameter values but have a mass range capped at that of the default Kroupa IMF: (0.03, 120) $\msun$. For the error-convolved power law, which has no defaults, we adopt $\alpha=2.35$ and $\sigma=0.5$ as in Figure \ref{fig:imfs}.
Testing was performed on one core of an AMD EPYC 7702 CPU within the University of Florida's HiPerGator cluster.
Results from performance testing are displayed in Table \ref{tab:samp_test}. All values correspond to the expected cost in time and memory to sample a $10^4$ $\msun$ cluster over the default mass range of each function; testing indicates that time for both methods and memory all scale linearly with cluster mass across the considered mass range. 
For the purposes of comparison with nonstellar use cases, the number of members in a star cluster also varies linearly with mass for the mass functions under consideration and is on the order of $10^4$ for a $10^4$ $\msun$ cluster.
\begin{deluxetable*}{c c c c c c c}
    \tabletypesize{\small}
    \tablewidth{\textwidth}
    \tablehead{\colhead{Quantity} &
    \colhead{Salpeter} &
    \colhead{Kroupa} &
    \colhead{Chabrier} &
    \colhead{Schechter} &
    \colhead{KK09\tablenotemark{a}} &
    \colhead{Padoan}}
    \tablecaption{Cost to draw $10^4$ \msun from all default IMF forms. Time is measured following both sampling methods, while approximate memory usage is measured for common functions. All quantities scale linearly with mass.}
    \label{tab:samp_test}
    \startdata
    Random sampling time\tablenotemark{b} (ms) & 0.5 & 1.1 & 1.9 & 12.0 & 3.1 & 0.9 \\
    Optimal sampling time (s) & 1.6 & 28.3 & 450.0 & 43.9 & 103.9 & 121.8 \\
    Memory usage (kB) & 77 & 185 & 114 & 764 & 160 & 65 \\
    \enddata
    \tablenotetext{a}{Parameter values are identical to Figure \ref{fig:imfs}.}
    \tablenotetext{b}{Values are averaged over all stop criteria.}
\end{deluxetable*}

\noindent\textbf{Random sampling.} Random sampling is generally fast regardless of the mass function or criterion. Basic power-law IMFs are the quickest to sample due to their low complexity, while more complicated IMFs with multiple components add to this time cost; for example, sampling the Kroupa IMF takes about twice as much time as a Salpeter IMF for each cluster mass. Changing the active stop criterion does not appreciably change the time to sample, as each algorithm broadly contains the same steps. 
Given the linear scaling of sampling time, \imf can be expected to provide sub-1 second performance for random sampling of clusters in most practical use cases.


\noindent\textbf{Optimal sampling.} Because optimal sampling is an iterative process requiring many more steps and operations than random sampling (see \S\ref{sec:sampling.2}), the time cost of optimally sampling a cluster is much higher than that of random sampling. 
The behavior of time cost with mass is roughly linear for all IMF forms, as in the case of random sampling. Optimal sampling is therefore relatively reasonable for lower-mass clusters, but the most massive clusters ($\approx10^5-10^6\,\msun$) are expensive to sample at scale. 
Chabrier IMFs are particularly slow; as implemented, the Chabrier mass functions rely on the base \texttt{scipy.stats} infrastructure to evaluate, thereby adding a time cost relative to custom \texttt{Distributions} with fewer internal operations.

\noindent\textbf{Memory.}
\imf formats its clusters as NumPy \texttt{array}s. 
Similar to time cost, memory usage is linear with cluster mass for each IMF form. IMFs that contain more low-mass stars lead to larger clusters, hence why the default Salpeter ($\mmin=0.3\,\msun$) and bottom-light default Padoan (see Figure \ref{fig:imfs}) functions generally take up less space than other functions ($\mmin=0.03\,\msun$).
Given the average expected size of sampled clusters, memory usage will likely not be a significant constraint for most use cases; however, it may become relevant when generating and storing many massive clusters given the accompanying large numbers of low-mass stars, particularly if the relevant mass function is bottom-heavy.

\section{Additional Functionality}\label{sec:func}

\subsection{Pre-/Protostellar populations}\label{sec:func.1}
The IMF, as a concept, describes the properties of fully formed stars. It is also closely related to the process of forming stars and is one of the observables that emerges from star formation theory. However, despite recent progress in determining the origin of the IMF, several aspects of the star formation process have few constraints placed on them, with the transition between the reservoirs of gas and dust that supply mass to forming stars to actual postformation stars remaining an area of considerable uncertainty \citep{hennebelle2024}. \imf provides classes that extend its functionality into the early lives of stars.

\subsubsection{PMFs}\label{sec:func.1.1}
The mass and luminosity functions of protostars, otherwise known as the protostellar mass/luminosity functions (PMF/PLF), have been proposed \citep[][M10/O11]{mckee2010,offner2011} and employed \citep{myers2014,hartmann2016} as observables that probe the phase of the star formation process between prestellar cores and stars. \imf implements PMFs as \texttt{MassFunctions} following the M10/O11 formalism, enabling performance of the same operations (integration, sampling\footnote{Note that sampling from a PMF provides protostellar masses, as opposed to the typical IMF sampling output of final stellar masses.}, etc.). 

PMFs are calculated by evaluating Equation 14 of M10:
\begin{equation}
    \psi_{\rm p}(m) = \frac{1}{\langle t_{\rm f}\rangle}\int_{m_{f,\,{\rm min}}}^{m_u} \xi(m_f)t_{\rm acc}(m_f,m)d m_f.
    \label{eq:pmf}
\end{equation}
$m$ is the current mass of a protostar, $m_f$ is the corresponding final mass, and $m_{f,\,{\rm min}}$ is max($m$,$\mmin$), the lowest possible final mass for a protostar with current mass $m$. $\xi$ is the IMF\footnote{In M10/O11, the IMF used is the log version, i.e. $dn/d\,ln\,m_f$, as opposed to $dn/dm_f$, the base definition in \imf. However, since the integral is done with respect to $d\,ln\,m_f$ in M10/O11, this implementation is functionally identical.}. $\langle t_{\rm f}\rangle$ is the IMF-weighted average of the time to form a star with final mass $m_f$ and $t_{\rm acc}$ is the characteristic accretion timescale for a protostar with mass $m$ and final mass $m_f$, equivalent to the current mass $m$ divided by the current accretion rate $\dot{m}$.

Calculating a PMF therefore requires the assumption of both a base IMF and an accretion history, i.e. a theory prescribing the rate of accretion onto protostars over time. M10 and O11 provide simple ``one-component" prescriptions based on the isothermal-sphere \citep[IS,][]{shu1977}, turbulent-core \citep[TC,][]{mckee2002,mckee2003}, and competitive-accretion \citep[CA,][]{bonnell1997,bonnell2001} theories of protostellar growth, as well as ``two-component" accretion rates that blend IS accretion with TC and CA.

One-component accretion rates follow the form
\begin{equation}
    \dot{m}=\dot{m}_1\left(\frac{m}{m_{\rm f}}\right)^j m_{\rm f}^{j_{\rm f}};
    \label{eq:mcmodel_1c}
\end{equation}
$m_{\rm f}$ is the final stellar mass, $j$ and $j_{\rm f}$ are derived from the accretion history, and $m_1$ is the final accretion rate for a 1 $\msun$ star. Two-component models follow the form
\begin{equation}
    \dot{m}=\dot{m}_{\rm IS}\left[1+\mathcal{R}_{\dot{m}}^2\left(\frac{m}{m_f}\right)^{2j} m_f^{2 j_f}\right]^{1/2};
    \label{eq:mcmodel_2c}
\end{equation}
$\dot{m}_{\rm IS}$ is the characteristic IS accretion rate and $\mathcal{R}_{\dot{m}}$ is the ratio between characteristic accretion rates for the blended histories (e.g. for two-component turbulent core, $\mathcal{R}_{\dot{m}}\equiv \dot{m}_{\rm TC}/\dot{m}_{\rm IS}$).

These accretion rate prescriptions may be further modified by tapering accretion or by assuming an exponentially increasing stellar birthrate, intended as a simple model for the accelerating star formation inferred in local star clusters by \citet{palla1999,palla2000}.
Tapered accretion is modeled by multiplying a protostar's accretion rate (Equations \eqref{eq:mcmodel_1c}, \eqref{eq:mcmodel_2c}) by a tapering factor, parameterized as $1-(t_m/t_f)^n$, where $n$ is a real number greater than 0. Accelerating star formation is modeled both by introducing a factor of $e^{-t_m/\tau}$ to the integrand in Equation \eqref{eq:pmf} and normalizing the PMF by $\tau\times\langle1-e^{t_f/\tau}\rangle$ in place of $\langle t_f\rangle$, where $\tau$ is a time constant. Both factors are applied inside the integral in Equation \eqref{eq:pmf}, as the current age of a protostar with mass $m$ $t_m$ is generically a function of current and final mass for accretion rates as modeled by M10/O11. These modifications may be made either separately or together; \imf calculates PMFs for each scenario. PMFs rely on interpolation for evaluation and sampling.

Much of the PMF formalism also applies to PLFs; in principle, they can also be implemented in the same way. \imf does not currently support PLFs, but support is planned; see Appendix \ref{ap:plf} for additional information.

\subsubsection{CMFs}\label{sec:func.1.2}
The core mass function (CMF) is the mass distribution of cores, which are dense condensations of gas and dust that comprise the gas reservoir from which stars form (though there remains ongoing debate on whether cores are well defined). The CMF has been postulated to be a precursor to the IMF based on its similar shape, making it a sought-after quantity in observations \citep[e.g.][]{motte1998,alves2007,konyves2015,motte2022}. \imf implements \texttt{MassFunctions} that calculate CMFs following the prescriptions of \citet{padoan2011} and \citet{hennebelle2008,hennebelle2009,hennebelle2013}, which derive CMF forms based on physical properties of the turbulent molecular clouds in which cores form.

\citet{padoan2011} CMFs are created by constructing a population of cores with randomly sampled masses, external densities, and ages. The mass associated with a core is the total mass it will assemble from the turbulent flow over its lifetime; these are drawn from an IMF as indicated in Section \ref{sec:sampling}. By default, this draw will be random sampling from an instance of a Salpeter IMF with the CMF's mass range, but the IMF and sampling method can be changed. External densities are randomly sampled from the standard lognormal gas density distribution, Equation 1 in \citet{padoan2011}\footnote{\imf employs the mass-weighted version of this equation to do this sampling, which is shifted to higher density instead of lower, in keeping with the framing of \citet{padoan2011}; see e.g. \citet{hopkins2013}.}. Ages are randomly sampled from a uniform distribution between 0 and the crossing time of the parent cloud, calculated from the input parameters. These core properties are then used to determine the current mass and evolutionary status of each core as it grows, which are used to construct the resulting CMF.

\begin{figure}
    \centering
    \includegraphics[width=0.75\textwidth]{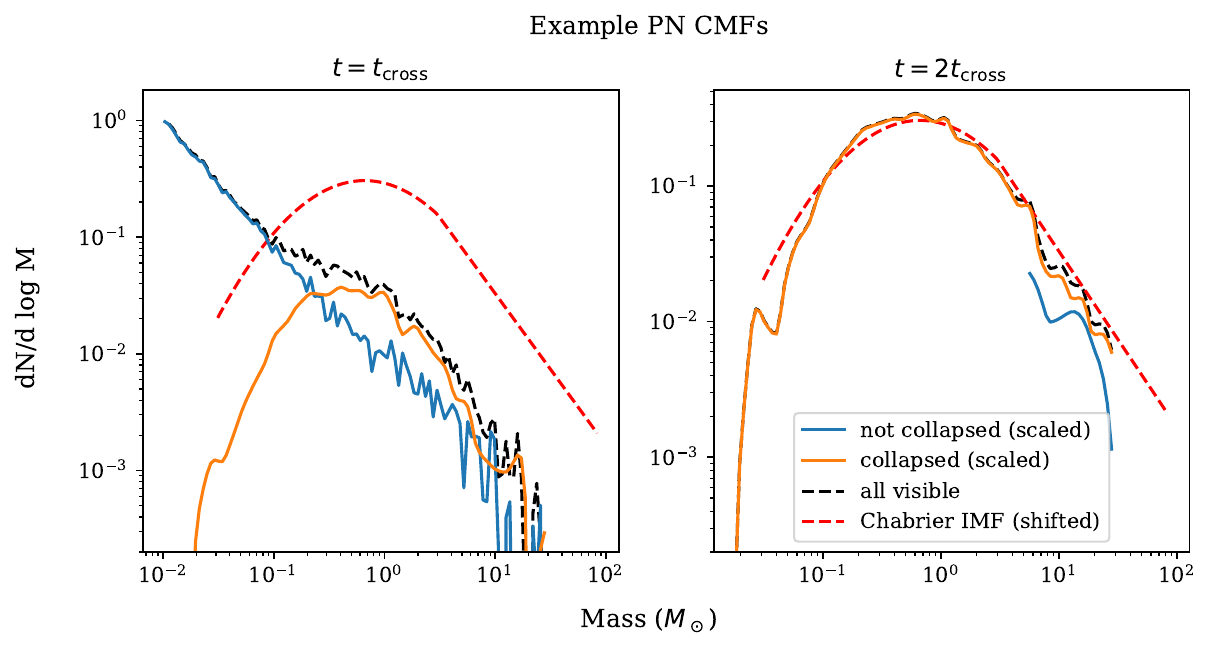}
    \caption{Mass functions derived using the approach of \citet{padoan2011}. \textbf{Left:} CMFs sampled at the crossing time of the parent cloud $t_{\rm cross}$ tracking visible cores. Transient and prestellar cores, those which have not yet collapsed, are plotted in blue while stellar cores, which have collapsed, are plotted in orange. The combined CMF is plotted in black; divided CMFs are scaled to match the total CMF. \textbf{Right:} The same functions, but now sampled at $t=2\times t_{\rm cross}$. The core population underlying both panels is the same. A \citet{chabrier2003a} IMF (\S\ref{sec:mfs.3})--shifted to higher mass by a factor of three to match the CMF--is plotted in red. At the later time, the total CMF is mainly comprised of collapsed cores and has attained a similar shape to the IMF.}
    \label{fig:pn_cmf}
\end{figure}

Cores composing the CMF are subdivided into three classes: ``transient" cores never reach their Bonnor-Ebert mass, ``prestellar" cores will reach their Bonnor-Ebert mass but are younger than the time it will take to reach that mass plus one freefall time (age $< t_{\rm BE} + t_{\rm ff}$), and ``stellar" cores will reach their Bonnor-Ebert mass and are older than that timescale. Bonnor-Ebert masses and freefall times are determined using external densities. Transient cores are only counted as ``visible" (i.e. potentially observable) while they are forming, while prestellar and stellar cores are always visible. By default, CMFs will include all cores that are visible and not stellar (i.e. forming transient cores and prestellar cores). However, \imf allows access to different core populations; for more details, see the documentation. Figure \ref{fig:pn_cmf} shows examples of CMFs generated from a fiducial core population using this framework.

Hennebelle \& Chabrier, by contrast, derive an analytic form for the distribution of core masses:
\begin{equation}
    \xi(M)\propto\frac{\bar{\rho}}{M}\frac{dR}{dM}\left[-\frac{d\delta}{dR}\,\exp(\delta)\,\mathcal{P}_R(\delta) +\int_{\delta_c}^\infty\exp(\delta)\frac{d{\mathcal{P}_R}}{dR}d\delta\right]
    \label{eq:hc_cmf}
\end{equation}
for core mass and radius $M$ and $R$, $\delta\equiv \ln(M/R^3)$, and gas density distribution $\mathcal{P}_R$. $\mathcal{P}_R$ is again taken to be lognormal, as in the Padoan \& Nordlund formalism; however, the variance of the distribution is taken to be a function of core size:
\begin{equation}
    \sigma^2(R) = \sigma_0^2\left[1-\left(\frac{R}{L_i}\right)^{n-3}\right]
    \label{eq:hc_sigma}
\end{equation}
where $\sigma_0^2=\ln(1+b^2\mathcal{M}^2)$ as in Section \ref{sec:mfs.6} (for forcing parameter $b$ and Mach number $M$), $n$ is the index of the 3D velocity power spectrum, and $L_i$ is the injection length for turbulence, assumed to be comparable to the $R$ of the parent cloud.

\begin{figure}
    \centering
    \includegraphics[width=0.6\textwidth]{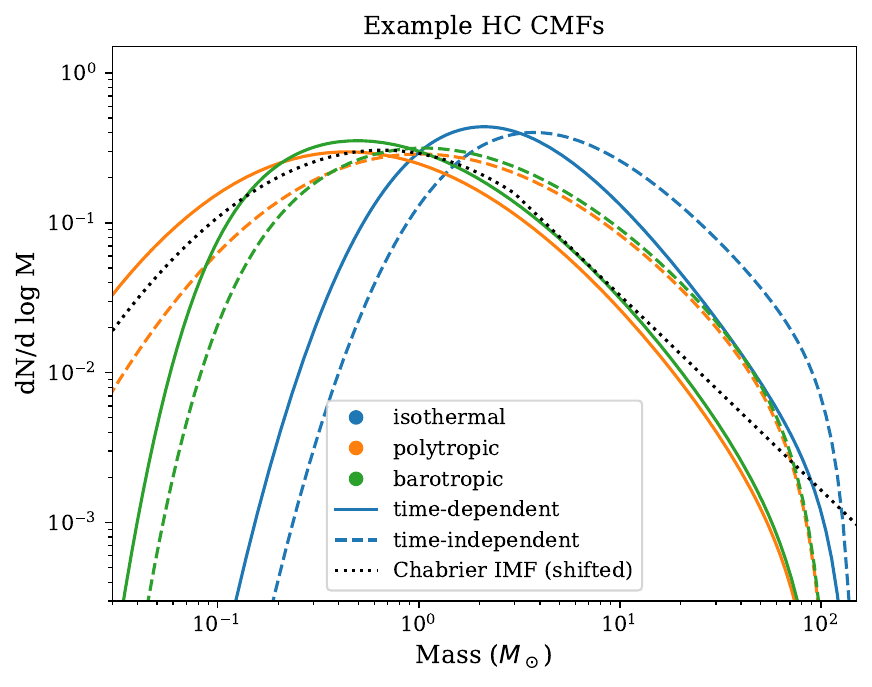}
    \caption{CMFs from \citet{hennebelle2008,hennebelle2009,hennebelle2013} plotted against the \citet{chabrier2003a} IMF shifted by a factor of 3 as in Figure \ref{fig:pn_cmf} \textit{(black, dotted)}. All CMFs assume default parameters, varying only in the assumed EOS and whether they are time-dependent \textit{(solid)} or independent \textit{(dashed)}.}
    \label{fig:hc_cmf}
\end{figure}

\imf implements Equation \eqref{eq:hc_cmf} as a \texttt{MassFunction} with an accompanying \texttt{Distribution} implementing this formalism. Across their series of papers, Hennebelle \& Chabrier consider multiple sources of support against collapse and equations of state (EOSs) for gas in the parent cloud that change the relationship between $M$ and $R$; all ``general case" forms assuming thermal and turbulent support are implemented. A function's EOS may be isothermal, polytropic (nonlinear), or barotropic (multicomponent), and additional support from a magnetic field may be toggled regardless of EOS. Further, each CMF has a time-independent and dependent version, which per \citet{hennebelle2013} differ by a factor $\propto\sqrt{M/R^3}$; all function instances provide access to both. Example HC CMFs are shown in Figure \ref{fig:hc_cmf}.

All CMFs employ interpolation for evaluation and sampling because their underlying mathematical forms are either complex or nonexistent.

\subsection{Luminosity calculations}\label{sec:func.2}
The IMF is commonly used to infer the properties of a zero-age stellar system from its integrated light. \imf provides tools to derive both the luminosity and Lyman continuum radiation from a sampled cluster, both as a whole and for individual members (although the calculation of Lyman continuum radiation is limited to stars with $\geq8\,\msun$). Lyman continuum is formatted as total luminosity $Q$, and \imf uses values corresponding to $Q_0$,
i.e. total luminosity from photons that ionize hydrogen. All returned values are logs, i.e. \imf returns $\log(L_{\rm tot})$ and $\log(Q_{\rm 0,\,tot})$ for a cluster of total mass $\mtot$.

By default, bolometric luminosities are interpolated from the stellar model grid of \citet[][E12]{ekstrom2012} and Lyman continuum luminosities are interpolated from the $Q_0$ values of Table 5 from \citet[][V96]{vacca1996}. Bolometric luminosities may also be interpolated from the same V96 table. Any values outside the native mass ranges of either grid ($0.8-64\,\msun$ for E12, $18.4-50\,\msun$ for V96) are extrapolated. For E12, luminosities are extrapolated assuming $L\propto M^{3.5}$ below its lower mass limit and $L\propto M^{1.35}$ above its upper mass limit. For V96, luminosities are extrapolated by continuing a line fit to the three values of $\log(m_\star)$ and $\log(L)/\log(Q_0)$ closest to either end of the defined mass range, with the exception of stars below $2\,\msun$. These low-mass stars do not emit Lyman continuum radiation, and their bolometric luminosities are scaled as $M^{0.23}$ from $0.03$ to $0.43\,\msun$ and $M^{4}$ from $0.43$ to $2\,\msun$. 

While \texttt{imf} can perform these very simple calculations for zero-age populations, it is not intended to be used for stellar population synthesis (SPS); however, it can be used to provide inputs for SPS codes.

\subsection{Visualization}\label{sec:func.3}
\begin{figure*}
    \centering
    \includegraphics[width=0.49\textwidth]{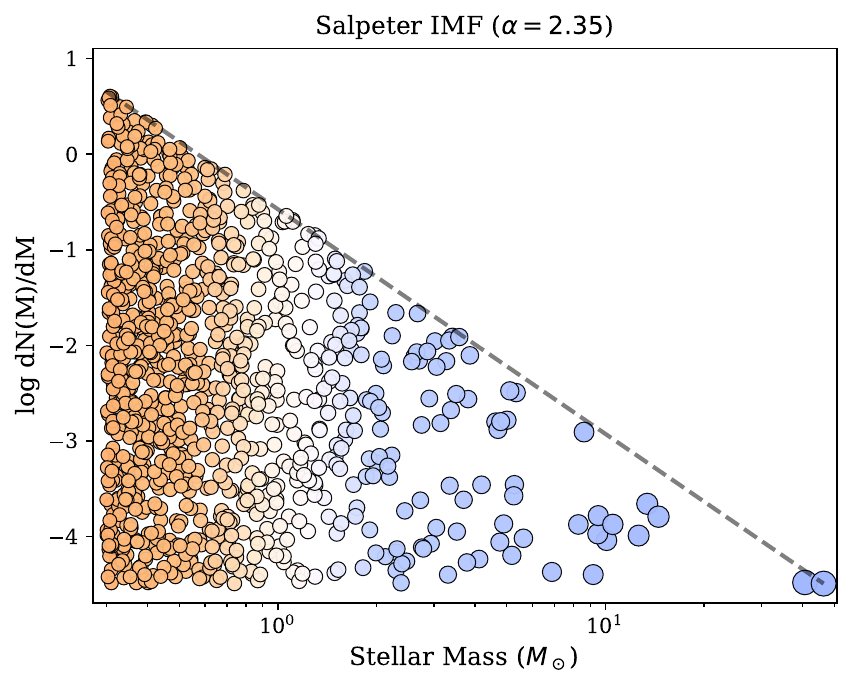}
    \includegraphics[width=0.49\textwidth]{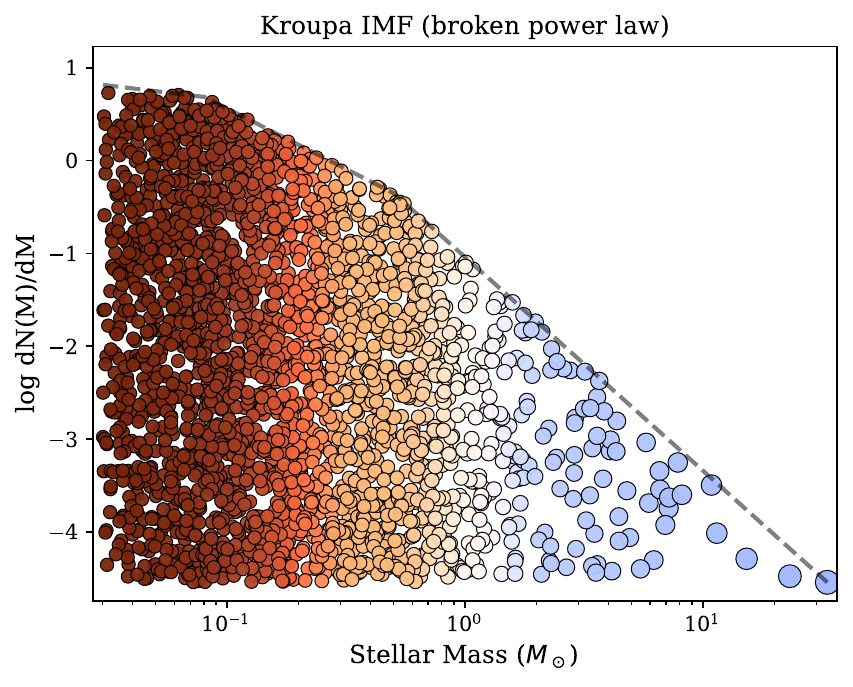}
    \includegraphics[width=0.49\textwidth]{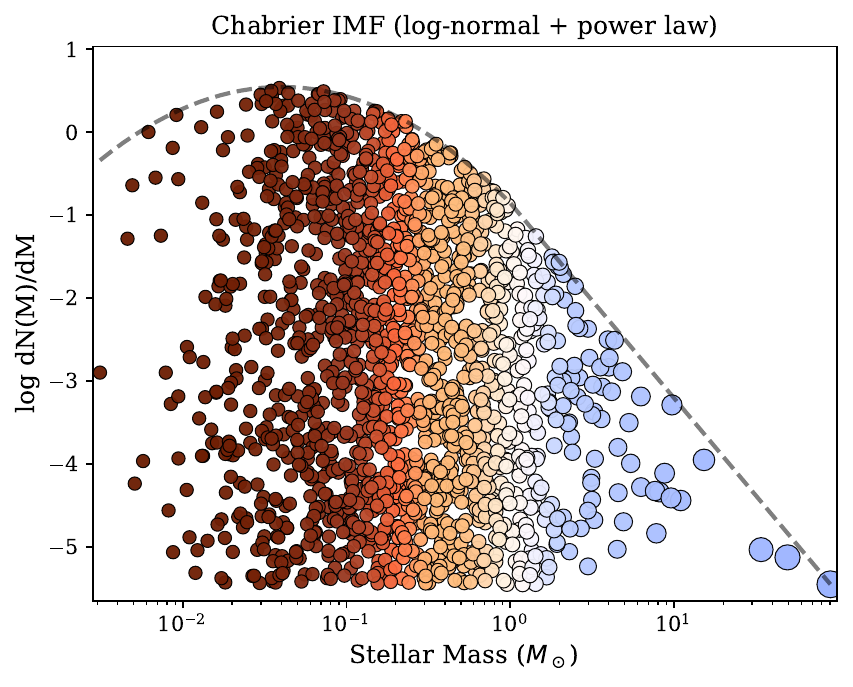}
    \caption{Visualizations of star clusters sampled with \imf's \texttt{make\_cluster} function according to the canonical Salpeter, Kroupa, and Chabrier IMFs (see \S\ref{sec:mfs}). Each cluster has $\mtot\approx1000\,\msun$. The color, size, and location of each data point is determined by the corresponding stellar mass. $y$-axis positions are randomly assigned to fill the space underneath the IMF.}
    \label{fig:base_functions}
\end{figure*}

\begin{figure*}
    \centering
    \includegraphics[width=0.49\textwidth]{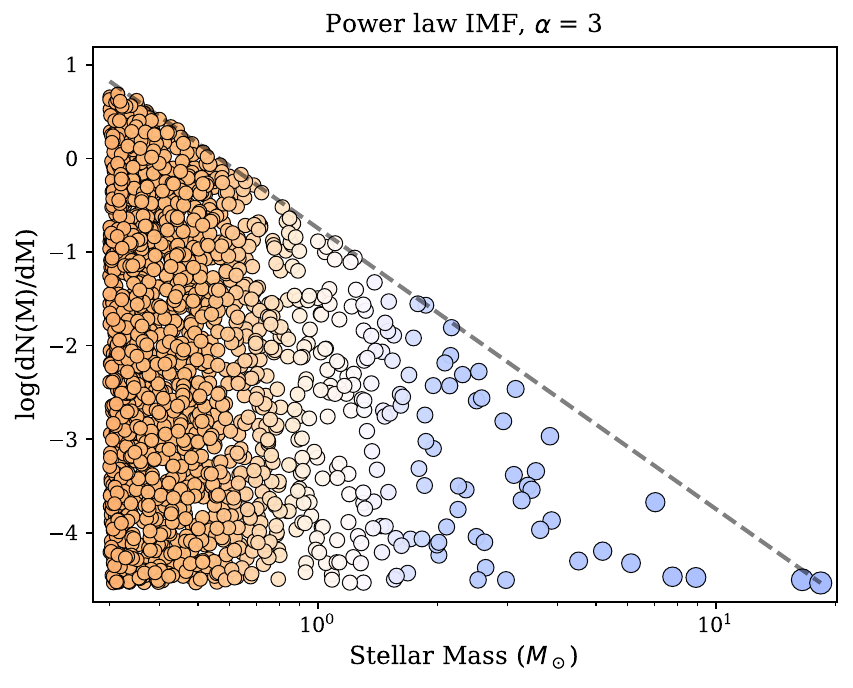}
    \includegraphics[width=0.49\textwidth]{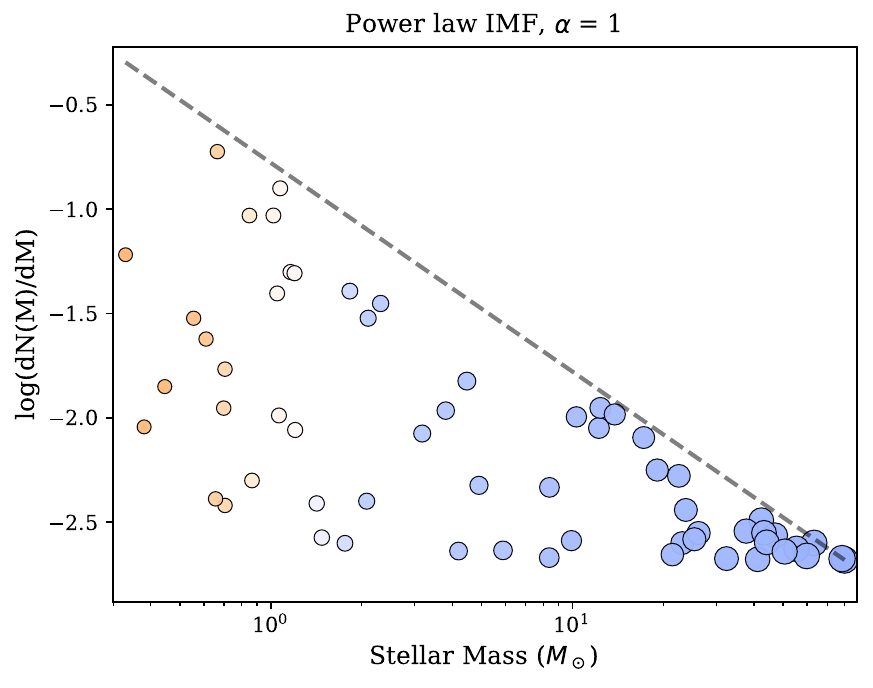}
    \caption{The same as Figure \ref{fig:base_functions}, but showing different power-law IMFs.}
    \label{fig:powerlaws}
\end{figure*}

\begin{figure*}
    \centering
    \includegraphics[width=0.49\textwidth]{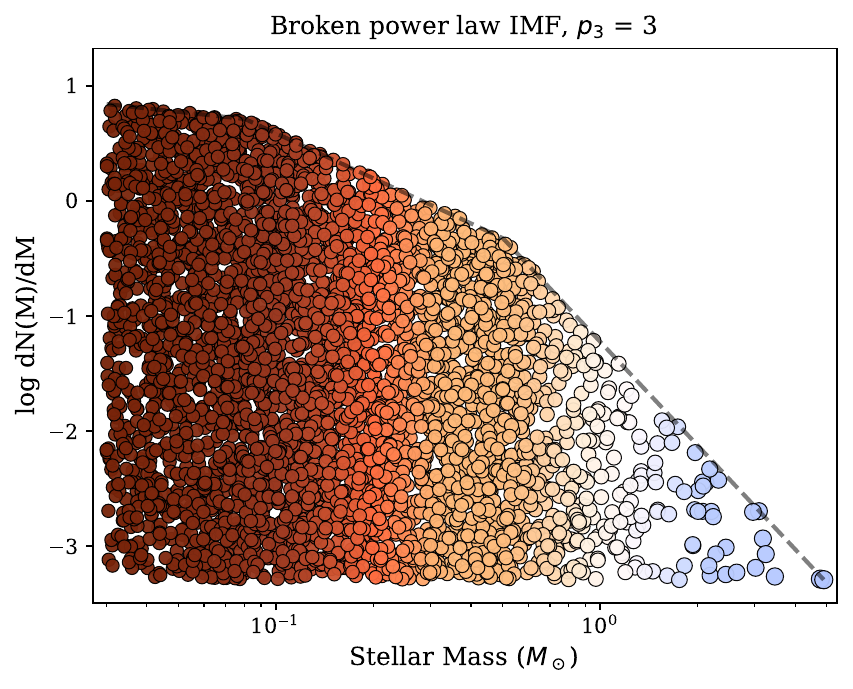}
    \includegraphics[width=0.49\textwidth]{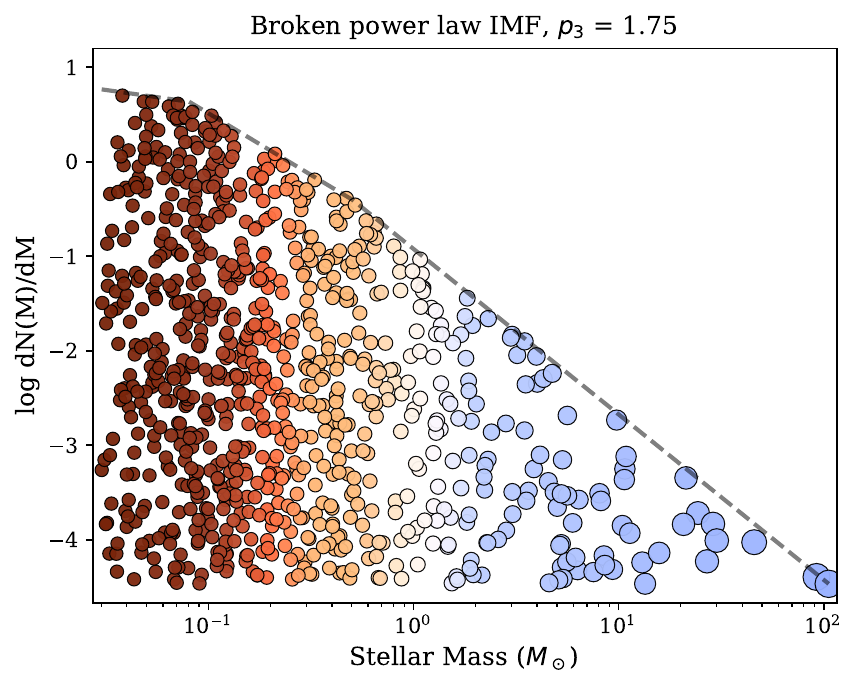}
    \caption{The same as Figure \ref{fig:base_functions}, but showing different broken power-law IMFs. Only the ``high-mass" (i.e. $m_\star > 0.5\,\msun$) slope is altered.}
    \label{fig:broken_powerlaws}
\end{figure*}

\imf provides functions that translate stars in sampled clusters into plottable data points, providing a more tactile way to display the results of sampling. Points have an $x$-axis location determined by mass and a $y$-axis position randomly sampled between the mass function's minimum value and value at a given $x$ such that the ensemble follows the mass function's shape. Each point is also given an associated color\footnote{\imf's colors originate from \href{vendian.org}{vendian.org}.} based on its mass; color corresponds roughly to main-sequence stellar type. Examples of this output are shown in Figures \ref{fig:base_functions}, \ref{fig:powerlaws}, and \ref{fig:broken_powerlaws}, which visualize clusters with total masses of approximately 1000 $\msun$ created by sampling common IMFs; these demonstrate both the function of \imf's visualization utilities and the impact of varying IMF forms and parameters on the resulting populations. \imf can also calculate a representative color for a cluster using the luminosity-weighted average color of its members; luminosities are calculated as described in Section \ref{sec:func.2}.

\section{Summary}\label{sec:end}
We have presented the \imf python package, a module designed to facilitate common uses for mass function models utilized across astronomy with a particular focus on the stellar IMF. This package enables sampling, integration, and basic characterization of populations spanning many object types across a wide range of mass functions. It also extends the capabilities of traditional IMF models into the pre- and protostellar phases of star formation, allowing for more robust analysis of the earliest life stages of stars. \imf is publicly available on the python package index as \texttt{initial\_mass\_function}\footnote{\href{https://pypi.org/project/initial-mass-function}{https://pypi.org/project/initial-mass-function}}. The version presented in this work is archived in Zenodo \citep{imfzenodo}; future development and releases can be tracked through the companion \href{https://github.com/keflavich/imf}{GitHub repository}.

\begin{acknowledgements}
    We thank Patrick Hennebelle for providing an IDL implementation of the Hennebelle \& Chabrier CMF, Jess Chellino for helpful discussions on the mathematics underlying the derivations of some functional forms, and the anonymous referee for useful suggestions. A.G. acknowledges support from the NSF under grants AST 2008101 and CAREER 2142300. S.K. acknowledges support from the Science \& Technology Facilities Council (STFC) grant ST/Y001001/1. The authors acknowledge \href{http://www.rc.ufl.edu}{University of Florida Research Computing} for providing computational resources and support that have contributed to the development of the software described in this publication.
\end{acknowledgements}

\software{Astropy \citep{astropy:2013, astropy:2018, astropy:2022}, astroquery \citep{astroquery}, Matplotlib \citep{matplotlib}, NumPy \citep{numpy}, SciPy \citep{scipy}.}

\appendix
\section{Protostellar luminosity functions}\label{ap:plf}
The formalism of the PLF is effectively the same as that of the PMF; they differ only in that the PLF $\Psi_{\rm p}(L)$ is obtained from the likelihood of a protostar existing at a particular luminosity given its final mass as opposed to that of existing at a particular mass. ``Luminosity" here refers to the total luminosity of a protostar, including both its intrinsic luminosity and luminosity originating from accretion. As such, the relevant equation is similar to Equation \eqref{eq:pmf}:
\begin{equation}
    \Psi_{\rm p}(L)=\int\Psi_{p2}(L,m_f)\,d\ln m_f.
\end{equation}
$\Psi_{p2}$ is the bivariate distribution of protostars according to luminosity and final mass; however, because luminosity is a function of current and final stellar mass for each of O11's modeled accretion histories, $\Psi_{p2}$ can be related to the PMF directly. O11 calculate PLFs by solving the following:
\begin{equation}
    \Psi_{\rm p}(L)=\frac{1}{\langle t_{\rm f}\rangle}\int_{m_{f,\,{\rm min}}}^{m_u} \xi(m_f)t_{\rm acc}(m_f,m) \left|\frac{\partial\ln m}{\partial\,\ln L}\right|d m_f,
    \label{eq:plf}
\end{equation}
which is the PMF--Equation \eqref{eq:pmf}--weighted by the derivative of $L(m,m_f)$ with respect to $m$.
\begin{figure}[ht]
    \centering
    \includegraphics[width=0.6\textwidth]{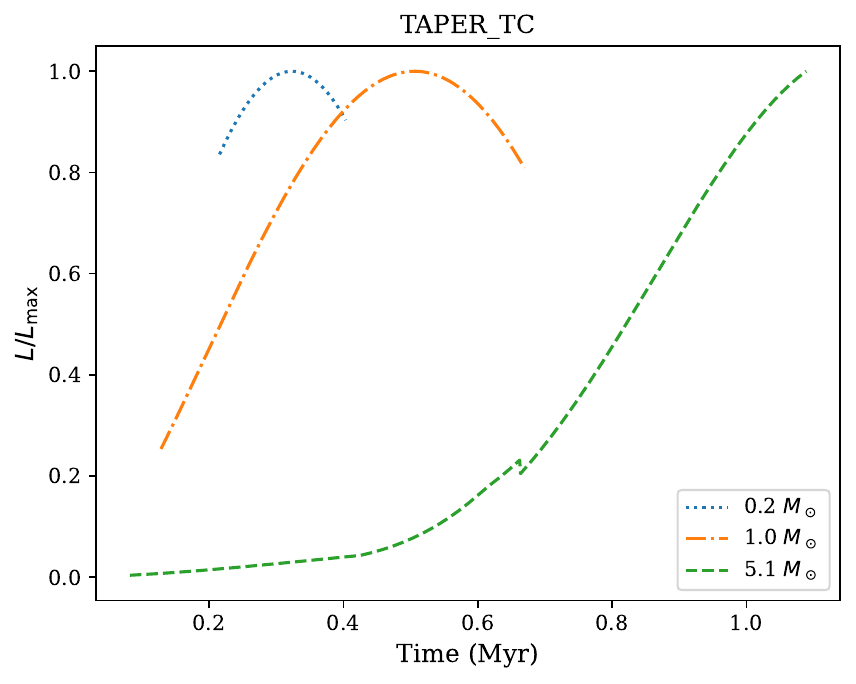}
    \caption{Time evolution of total (intrinsic + accretion) luminosity for protostars of low to intermediate mass following a tapered turbulent core accretion history as described in M10. Values are derived using a version of the protostellar evolution code of \citet{klassen2012} altered to model other accretion histories \citep[done as part of][]{richardson2025}.
    Tracks begin when the protostar is initialized and end when accretion stops. Luminosities are plotted as a fraction of the maximum attained by each protostar for display purposes.}
    \label{fig:plf}
\end{figure}

Calculating a PLF using Equation \eqref{eq:plf} requires the ability to map from current/final protostellar mass $m/m_f$ to total luminosity $L$ in order to calculate the gradient of $L$, and likewise from $L$ and $m_f$ to $m$ in order to determine a protostar's accretion time $t_{\rm acc}$ (a function of $m$, not $L$). However, these mappings do not have simple analytic forms, nor are they generically expected to be bijective, as the evolution of a protostar's luminosity is not required to be monotonic (and often is not depending on stellar structure or assumed accretion history). Figure \ref{fig:plf} demonstrates common ways in which the mapping of $L$ to $m$ may be complicated. Low-mass protostars whose luminosity is dominated by accretion can exhibit the same $L$ at multiple $m$ if mass-driven growth in intrinsic luminosity is offset by the waning accretion luminosity occurring in tapered accretion histories; meanwhile, protostars that become massive enough to initiate deuterium burning and enter the Henyey track while accreting experience a discontinuity in radius (and therefore luminosity) due to rapid swelling caused by the switch to a radiative core \citep[see][]{offner2009}.

Calculating a PLF consequently requires additional algorithmic considerations that are difficult to implement without making simplifying assumptions that may not be generally applicable; for example, assuming that accretion luminosity is dominant makes the PLF easier to work with analytically but is also only applicable to low-mass stars. \imf therefore does not currently calculate PLFs; this functionality is planned for a future update, but is dependent on the tractability of these foundational problems.

\bibliography{bib}{}
\bibliographystyle{aasjournalv7}

\end{document}